\documentclass[twocolumn,twoside,slac_two]{revtex4}
\usepackage{graphicx}
\usepackage{fancyhdr}
\begin{document}

\title{MAGIC and Multi-Wavelength Observations of three HBLs in 2008}

%

\author{S. R\"ugamer}
\affiliation{Institut f\"ur Theoretische Physik und Astrophysik, Universit\"at W\"urzburg, D-97074 W\"urzburg, Germany}
\author{E. Angelakis}
\affiliation{Max-Planck-Institut f\"ur Radioastronomie, D-53121 Bonn, Germany}
\author{D. Bastieri}
\affiliation{Universit\`a di Padova and INFN, I-35131 Padova, Italy}
\author{D. Dorner}
\affiliation{ISDC Data Center for Astrophysics, CH-1290 Versoix, Switzerland}
\author{Y. Y. Kovalev, K. V. Sokolovsky}
\affiliation{Max-Planck-Institut f\"ur Radioastronomie, D-53121 Bonn, Germany\\Astro Space Center of Lebedev Physical Institute, 117997 Moscow, Russia}
\author{A. L\"ahteenm\"aki}
\affiliation{Aalto University Mets\"ahovi Radio Observatory, FI-02540 Kylm\"al\"a, Finland}
\author{E. Lindfors, R. Reinthal}
\affiliation{Tuorla Observatory, University of Turku, FI-21500 Piikki\"o, Finland}
\author{C. Pittori}
\affiliation{Agenzia Spaziale Italiana (ASI) Science Data Center, I-00044 Frascati, Italy}
\author{A. Stamerra}
\affiliation{Universit\`a  di Siena, and INFN Pisa, I-53100 Siena, Italy}
\author{H. Ungerechts}
\affiliation{Instituto de Radio Astronoma Milim\'{e}trica, E-18012 Granada, Spain\\on behalf of the MAGIC Collaboration, the F-Gamma program and the Fermi LAT, RATAN, AGILE and IRAM team}

\begin{abstract}
The high-frequency peaked blazars and known TeV emitters 1ES\,1011+496, Mrk\,180 and 1ES\,2344+514 have been observed in the course of multi-wavelength campaigns in 2008, covering the frequency bands from radio up to TeV energies. For all three sources, these coordinated observations represent the first of their kind. We will present and discuss the campaigns, resulting light curves and spectral energy distributions.
\end{abstract}

\maketitle

\thispagestyle{fancy}


\section{Introduction}
The extragalactic Very High Energy (VHE, $\gtrsim$ 100\,GeV) gamma-ray sky is dominated by blazars, active galactic nuclei whose non-thermal emission is emanating from a relativistic plasma jet closely aligned to our line of sight \cite{blazars}. Due to beaming effects, these sources are the brightest and most variable extragalactic gamma-ray emitters. Their spectral energy distribution (SED) shows two pronounced peaks (in a double-logarithmic energy flux vs.\ frequency representation), one around optical to X-ray energies, the other one located at gamma-ray energies. According to the position of the first peak, BL Lacertae objects (blazars without strong spectral lines) are categorised as high-frequency peaked BL Lac objects (HBLs) if the peak energy falls within the UV to soft X-ray regime (e.g.\ \cite{HBL}). The first peak is thought to be produced by synchrotron radiation from relativistic electrons accelerated in the jets of the blazar, whereas the origin of the second peak is still a matter of debate. Most of the measured SEDs can be described by the widely used Synchrotron Self-Compton models (e.g.\ \cite{SSC1, SSC2}), though some sources require additional emission components like external radiation fields (e.g.\ \cite{EC1, EC2}). Also hadronic models (e.g.\ \cite{PIC1, PIC2}) are successfully applied. Due to lack of constraining data, it is until now not possible to discriminate between the different kind of models, which also makes it difficult to determine the physics at work in these sources.

Blazars show variability in flux as well as spectral shape from radio to VHE frequencies at timescales down to minutes (e.g.\ \cite{MAGIC_Mrk501}). Consequently, simultaneous observations at all involved frequency bands are necessary to retrieve a reliable SED. The complexity and amount of inter-collaborative efforts of such observation campaigns as well as the low sensitivity of the first generation of gamma-ray instruments made these campaigns quite rare. At the time of 2008, only a handful of sources had been studied in simultaneous multi-wavelength (MW) campaigns, dominantly centered on the brightest objects or high flux states. Also just since end of 2007 and mid of 2008, respectively, the $AGILE$ and $Fermi$ satellites are available which cover for the first time since EGRET (mission end: 2000) the crucial high energy (HE, $\gtrsim$ 100 MeV) regime.

The campaigns described in the following were organised intentionally regardless of the flux state of the source to obtain an unbiased sample of simultaneously measured SEDs for objects hardly studied in MW campaigns until now. The MW light curve plots of these campaign can be found in \cite{1011_ICRCproceedings} and \cite{Mrk180_2344_ICRCproceedings}.

\section{Observational Targets}
All three objects of these campaigns, 1ES\,1011+496 (redshift $z = 0.212$), Mrk\,180 ($z = 0.046$) and 1ES\,2344+514 ($z = 0.044$), are HBLs detected at TeV energies. The first two objects were discovered at VHE by MAGIC in 2007 \cite{1011_MAGIC} and 2006 \cite{Mrk180_MAGIC}, respectively, following an optical high state. 1ES\,2344+514 on the other hand is a known VHE emitter since 1998 \cite{2344_Whipple}. For an overview of historical observations of the three sources at different wavelength see \cite{1011_ICRCproceedings} and \cite{Mrk180_2344_ICRCproceedings}, respectively, as well as the references mentioned above. None of the objects have been observed in simultaneous MW campaigns from radio to TeV energies until now.

\section{Observation Campaigns and Results}
\subsection{1ES\,1011+496}
The observations have been conducted from March till June 2008. MAGIC observed during 25 nights as close to the two $AGILE$ observation windows (spanning 14 and 10 days, respectively) as possible. The achieved simultaneous coverage was limited by adverse meteorological conditions like cloud coverage or calima at the MAGIC site. Optical observations were provided by the KVA telescope, which is operated concurrently with the MAGIC telescope. Additionally, $Swift$ ToO observations were carried out for 10 days during the campaign, and the radio telescope Mets\"ahovi observed on 2 nights at 37\,GHz.

The preliminary analysis shows that MAGIC detected the source clearly with a significance of $> 7 \sigma$ during the campaign. The light curve was consistent with a constant flux, though for one night the flux was $> 2 \sigma$ above the average flux. The average spectrum derived from the observations showed a flux comparable with the one during the MAGIC detection and compatible but slightly harder spectral index.

At X-rays, $Swift$ XRT detected a small flare, where the flux rose by $\sim 40 \%$ within 5 days and subsequently decreased to half of the flare peak flux after 6 days. As coverage and hence the flux baseline before and after this flare is missing, the overall flare rise and fall times could not be evaluated. The changes of the XRT spectral indices during that flare were clearly correlated with the changes of the X-ray fluxes.

In the optical regime, significant variability has been found in the R band. The optical flux rose from relatively low levels by $\sim 50 \%$ towards the end of the campaign. Measurements by KVA in the V and B band from end of April till beginning of May as well as in the $Swift$ UVOT filters followed in general the trend in the R band, showing only moderate flux changes in this time period. A correlation of the R band behaviour with other frequency bands cannot be investigated due to missing coverage at other wavelengths.

Within the given time windows, Mets\"ahovi and $AGILE$ did not detect the source.

\subsection{Mrk\,180}
Most of the Mrk\,180 observations were conducted in two separate windows, one around beginning of May, the other one from end of October till beginning of December 2008. MAGIC observed on 24 nights, mostly during moderate moonlight or twilight, accompanied by dense KVA monitoring. Two $AGILE$ windows were covered, the first around May spanning 10 days, the second in November (30 days). For both windows, $Swift$ ToO pointings could be arranged for altogether 21 nights. At radio frequencies, several snapshot observations by Effelsberg, RATAN-600, Mets\"ahovi and IRAM were conducted. Additionally, the $Fermi$ satellite was already in orbit during the second observation window.

The MAGIC threshold for this observations was strongly increased due to the high zenith angle of the source and the increased background noise due to moonlight and twilight. Additional cloud coverage and calima rendered only $\sim 7$ hours out of $> 15$ hours of data taken usable for further analysis. The analysis results will be presented soon.

$Swift$ XRT detected significant variability in both time windows. In the first window, the flux increased towards the end by more than a factor of 2, whereas in the second window, a huge flare with a flux increase by a factor of 9 was detected. This represents the strongest flare and highest flux ever measured for this object at X-rays. Within 6 days, the flux was declining to only $\sim 35 \%$ of the peak flux and reaching the baseline level again after another 11 days. The strong signal should allow to detect intra-night variability, but for the three nights with the highest fluxes, significant variability was not present investigating time bins of $200$\,s. Also in this case, the XRT flux and spectral index were significantly correlated.

A HE flare, lasting for $\sim 17$ days, could be detected by $Fermi$-LAT, which ended $\sim 15$ hours before the highest flux at X-rays was reached. Due to missing X-ray data during the HE flare, a direct correlation cannot be excluded nor confirmed.

The KVA R band flux was overall higher during the second window than during the first window and showing in both windows decreasing trends. But the optical light curve didn't show correlations on daily scale with the observed X-ray flares. Also a correlation between the $Swift$ UVOT and XRT measurements is not apparent.

At radio frequencies $< 50$\,GHz, flux variability is present over timescales of month; shorter timescales cannot be tested due to the low sampling at these frequencies. The better sampled IRAM light curve showed a constant flux over months at 86\,GHz but a rise on the last day of observations (30 days before the start of the $Fermi$-LAT flare).

Mets\"ahovi and $AGILE$ did not detect the source during the campaign.

\subsection{1ES\,2344+514}
For 1ES\,2344+514, the same wavelength coverage was achieved as for Mrk\,180 but additional VLBA observations. The campaign took place between September 2008 and January 2009. MAGIC observed the source on 17 nights, collecting a total of $\sim 21$ hours good quality data. $Swift$ observations were conducted on 21 days.

MAGIC did not detect the source with $> 5 \sigma$. Considering that 1ES\,2344+514 is a well-established VHE source, the rather long observation time of $> 20$ hours and the good data quality, a spectrum and light curve was derived nevertheless. The low significance of the signal should be kept in mind, though. The source was found on a flux level slightly below the one found by MAGIC in 2005 with a spectral index slightly harder than during the MAGIC detection, but consistent with that value within the error bars. The light curve did not show any hint of variability.

The optical R band light curve measured by KVA was also consistent with a constant flux, whereas at X-rays, $Swift$ XRT found significant variability. The flux increased by $\sim 50 \%$ within 2 days and dropped to about half the peak flux value within 8 days. Spectral variability was present during the flare, and again the spectral index was clearly correlated with the flux. The overall flux level was among the lowest ever measured at X-rays for 1ES\,2344+514.

At radio wavelength, the source showed variability on the sampling timescale of $\sim 4$ weeks at frequencies below 50\,GHz. The IRAM light curve did not show a hint of variability throughout the observations. Correlated behaviour with other frequency bands could not be investigated due to the different sampling.

Mets\"ahovi detected the source three month before the campaign, but not during the coordinated observations. This excludes major flares to have happened at 37\,GHz during the campaign. Also $AGILE$ and $Fermi$-LAT did not detect the source in the given time windows.

\section{Discussion and Conclusions}
For each source two simultaneous SEDs were constructed. These data sets have been chosen according to the X-ray flux state with the criteria of (1) a significant flux difference between the two data sets and (2) having the best possible simultaneous coverage at all wavelengths. The SEDs were modelled using a one-zone SSC model \cite{Tavecchio03} as well as a self-consistent two-zone SSC model \cite{Weidinger}. Note that the data collection and modelling is still ongoing.

\begin{figure}[t!p]
\centering
\includegraphics[clip, width=3.2in]{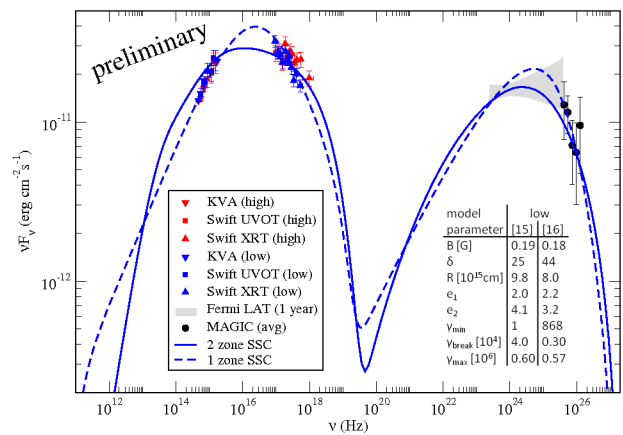}
\caption{SED of 1ES\,1011+496 resulting from this campaign. The KVA and UVOT data are host-galaxy corrected, the UVOT data additionally de-reddened. The grey bow-tie is deduced from the LAT 1-year Catalog, the VHE points represent the average MAGIC spectrum (already corrected for EBL effects using \cite{Kneiske_lowerlimit}).}
\label{fig:SED_1011}
\end{figure}

Due to the rather small difference in flux in the case of 1ES\,1011+496, only one model fit was applied to the two data sets (see Figure \ref{fig:SED_1011}). The resulting model parameters are typical for HBLs but the high Doppler factor and $\gamma_{min}$ of \cite{Weidinger}, which may be attributed to the preliminary nature of the modelling and will be refined in the near future. Comparing the MAGIC discovery spectrum with the one derived here and taking into account that the Fermi bow-tie shown in the SED fits the VHE data rather well though being taken month after the actual campaign, 1ES\,1011+496 seems to be a rather constant gamma-ray emitter. More observations are of course necessary to confirm that interpretation. The synchrotron component in the optical and X-rays shows significant variability instead. In \cite{1011_MAGIC}, non-contemporaneous X-ray data, at a factor 10 lower in flux than the ones measured here simultaneously, was used for modelling the SED. The results led to the conclusion that 1ES\,1011+496 would be a Compton dominated source, unlike most of the other HBLs. The results derived here show instead that also this source seems to by synchrotron dominated.

\begin{figure}[t!p]
\centering
\includegraphics[clip, width=3.2in]{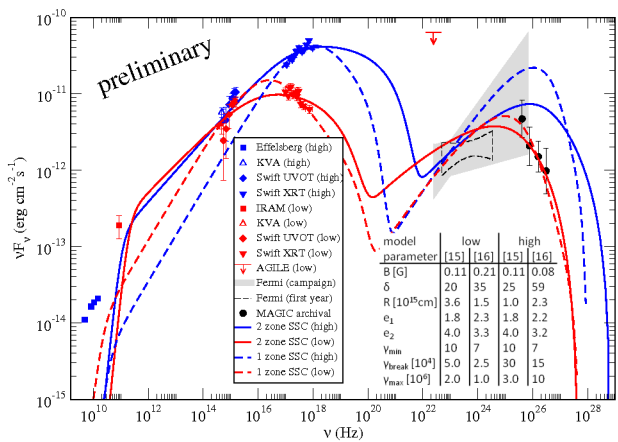}
\caption{SEDs of Mrk\,180 from the campaign presented here. The KVA data are host-galaxy corrected, the UVOT data are de-reddened and the thermal contribution has been subtracted. At high gamma-ray energies, a $95 \%$ c.l.\ upper limit derived by $AGILE$ is shown together with $Fermi$ bow-ties deduced from the LAT 1-year Catalog (dashed line) and from the MAGIC-simultaneous period (grey shaded). The shown MAGIC data points are taken from \cite{Mrk180_MAGIC}.}
\label{fig:SED_Mrk180}
\end{figure}

The simultaneous SED of the Mrk\,180 low X-ray flux state could be modelled well with both models (see Figure \ref{fig:SED_Mrk180}), also yielding rather standard parameters. On the contrary, the rather steep X-ray spectrum and high optical flux made it difficult for both models to describe the high flux data. Either they have to assume a different spectral shape in X-rays, or underestimate the optical flux. Despite these modifications, $\gamma_{break}$ and $\delta$ are high for both models. One solution to explain this discrepancy would be to assume that the optical and X-ray emission originates in different regions of the jet. Further modelling is in progress. The unprecedented high X-ray flux state of Mrk\,180 was accompanied by a strong shift of the synchrotron peak, from $\lesssim 0.5$\,keV to $\gtrsim 5$\,keV, making also Mrk\,180 an 'extreme blazar' (\cite{extreme_blazars}) candidate.
\begin{figure}[t!p]
\centering
\includegraphics[clip, width=3.2in]{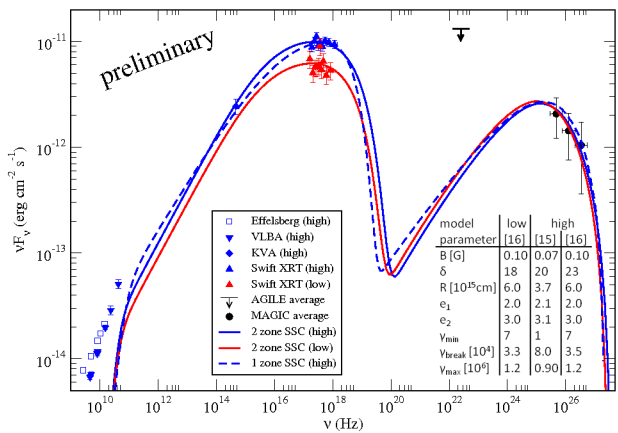}
\caption{SEDs of 1ES\,2344+514. The KVA data are corrected for host galaxy effects. The $AGILE$ upper limit shown at high gamma-ray energies has a c.l.\ of $95 \%$. The black data points denote the average spectrum measured by MAGIC, corrected for EBL absorption by \cite{Kneiske_lowerlimit}.}
\label{fig:SED_2344}
\end{figure}

1ES\,2344+514 was observed in one of the lowest flux states ever at VHE, X-rays and optical wavelengths. The SEDs in low and high flux state could nevertheless be described well by the applied models (see Figure \ref{fig:SED_2344}), resulting in typical HBL parameter values. Consequently we either have not yet observed the 'quiescent state' of the source, or this flux state is not characterised by model parameters different from the standard ones published up to now. From the VLBA observations, upper limits on the size and magnetic field of the radio emitting region could be derived \cite{2344_VLBA} which are in agreement with the model parameters found for the blazar emission zone here.


\bigskip 
\begin{acknowledgments}
\begin{scriptsize}
The MAGIC collaboration would like to thank the Instituto de Astrof\'{\i}sica de Canarias for the excellent working conditions at the Observatorio del Roque de los Muchachos in La Palma. The support of the German BMBF and MPG, the Italian INFN, the Swiss National Fund SNF, and the Spanish MICINN is gratefully acknowledged. This work was also supported by the Marie Curie program, by the CPAN CSD2007-00042 and MultiDark CSD2009-00064 projects of the Spanish Consolider-Ingenio 2010 programme, by grant DO02-353 of the Bulgarian NSF, by grant 127740 of the Academy of Finland, by the YIP of the Helmholtz Gemeinschaft, by the DFG Cluster of Excellence ``Origin and Structure of the Universe'', by the DFG Collaborative Research Centers SFB823/C4 and SFB876/C3, and by the Polish MNiSzW grant 745/N-HESS-MAGIC/2010/0.

The $Fermi$ LAT Collaboration acknowledges support from a number of agencies and institutes for both development and the operation of the LAT as well as scientific data analysis. These include NASA and DOE in the United States, CEA/Irfu and IN2P3/CNRS in France, ASI and INFN in Italy, MEXT, KEK, and JAXA in Japan, and the K.\ A.\ Wallenberg Foundation, the Swedish Research Council and the National Space Board in Sweden. Additional support from INAF in Italy and CNES in France for science analysis during the operations phase is also gratefully acknowledged.

The AGILE Mission is funded by the Italian Space Agency (ASI) with scientific and programmatic participation by the Italian Institute of Astrophysics (INAF) and the Italian Institute of Nuclear Physics (INFN).

We gratefully acknowledge N.\ Gehrels for approving this set of ToOs and the entire $Swift$ team, the duty scientists and science planners for the dedicated support, making these observations possible.

\end{scriptsize}
\end{acknowledgments}

\bigskip 

\begin{thebibliography}{99} 
\begin{scriptsize}
\bibitem{blazars} Urry, C.\ M., \& Padovani, P.,1995, PASP 107, 803
\bibitem{HBL} Padovani, P., \& Giommi, P., 1995, ApJ 444, 567
\bibitem{SSC1} Maraschi, L., Ghisellini, G., Celotti, A., 1992, ApJ 397, L5
\bibitem{SSC2} Costamante, L., \& Ghisellini, G., 2002, A\&A 384, 56
\bibitem{EC1} Dermer, C.\ D., \& Schlickeiser, R., 1993, ApJ 416, 458
\bibitem{EC2} Ghisellini, G., Tavecchio, F., Chiaberge, M., 2005, A\&A 432, 401
\bibitem{PIC1} Mannheim, K., 1993, A\&A 269, 67
\bibitem{PIC2} M\"ucke, A., et al., 2003, Astropart.\ Phys.\ 18, 593
\bibitem{MAGIC_Mrk501} Albert, J., et al. (MAGIC Collab.), 2007, ApJ 669, 862
\bibitem{1011_ICRCproceedings} Reinthal, R., et al., 2011, Proc. of the 32nd ICRC, Beijing, China, arXiv:1109.6504 [astro-ph]
\bibitem{Mrk180_2344_ICRCproceedings} R\"ugamer, S., et al., 2011, Proc. of the 32nd ICRC, Beijing, China, arXiv:1109.6808 [astro-ph]
\bibitem{1011_MAGIC} Albert, J., et al.\ (MAGIC Collab.), 2007, ApJ 667L, 21
\bibitem{Mrk180_MAGIC} Albert, J., et al.\ (MAGIC Collab.), 2006, ApJ 648L, 105
\bibitem{2344_Whipple} Catanese, M., et al., 1998, ApJ 501, 616
\bibitem{Tavecchio03} Maraschi, L., \& Tavecchio, F., 2003, ApJ 593, 667
\bibitem{Weidinger} Weidinger, M., \& Spanier, F., 2010, A\&A 515, A18
\bibitem{Kneiske_lowerlimit} Kneiske, T.\ M., \& Dole, H., 2010, A\&A 515, A19
\bibitem{extreme_blazars} Costamante, L., et al., 2001, A\&A 371, 512
\bibitem{2344_VLBA} Sokolovsky, K.\ V., et al., 2010, in Proc.\ of the Workshop ``Fermi meets Jansky - AGN in Radio and Gamma-Rays``, Bonn, Germany, arXiv:1006.3084 [astro-ph]

\end{scriptsize}

\end{thebibliography}

\end{document}